# Retrospective Estimation of the Quality of Intensity-Modulated Radiotherapy Plans for Lung Cancer

Jihye Koo

*Radiologic Science, Korea University, Seoul, Korea 136-703*

Myonggeun Yoon

*Department of Bio-Convergence Engineering, Korea University, Seoul, Korea 136-703*

Weon Kyu Chung and Dong Wook Kim

*Department of Radiation Oncology, Kyung Hee University Hospital at Gangdong,*

*Seoul, Korea 134-727*

This study estimated the planning quality of intensity-modulated radiotherapy in 42 lung cancer cases to provide preliminary data for the development of a planning quality assurance algorithm. Organs in or near the thoracic cavity (ipsilateral lung, contralateral lung, heart, liver, esophagus, spinal cord, and bronchus) were selected as organs at risk (OARs). Radiotherapy plans were compared using the conformity index (CI), coverage index (CVI), and homogeneity index (HI) of the planning target volume (PTV), OAR-PTV distance and OAR-PTV overlap volume, and the $V10_{Gy}$, $V20_{Gy}$, and equivalent uniform dose (EUD) of the OARs. The CI, CVI, and HI of the PTV were 0.54–0.89 (0.77 ± 0.08), 0.90–1.00 (0.98 ± 0.02), and 0.11–0.41, (0.15 ± 0.05), respectively. The mean EUDs ($V10_{Gy}$, $V20_{Gy}$) of the ipsilateral lung, contralateral lung, esophagus, cord, liver, heart, and bronchus were 8.07 Gy (28.06, 13.17), 2.59 Gy (6.53, 1.18), 7.02 Gy (26.17, 12.32), 3.56 Gy (13.56, 4.48), 0.72 Gy (2.15, 0.91), 5.14 Gy (19.68, 8.62), and 10.56 Gy (36.08, 19.79), respectively. EUDs tended

to decrease as OAR-PTV distance increased and OAR-PTV overlap volume decreased. Because the plans in this study were from a single department, relatively few people were involved in treatment planning. Differences in treatment results for a given patient would be much more pronounced if many departments were involved.

Email: joocheck@gmail.com, dwkim@khnmc.or.kr,

Fax: +82-2-440-7393

## I. INTRODUCTION

Lung cancer is a high-risk disease that accounts for the majority of deaths from cancer according to a study performed in the National Cancer Information Center in Korea in 2013 [1]. As reported in 2012, it is the fourth most common form of cancer after thyroid, stomach, and colon cancer [2]. Surgical resection is the standard treatment for primary lung cancer, but the early detection rate is only 20–25%. Treatment of advanced lung cancer generally includes chemotherapy and radiotherapy (RT) along with surgical resection, but is dependent on the patient's status. Concurrent chemo-radiation therapy has been shown to considerably increase survival rates in advanced lung cancer [3].

Although RT is an efficient cancer treatment method, complications such as pneumonitis are inevitable in most cases [4–5]. Thus, it is important to minimize normal tissue irradiation by optimizing treatment plans, reducing set-up margins via precise patient positioning, and ensuring accurate mechanical performance. To achieve this goal, there have been continuous efforts to develop quality assurance (QA) protocols to assure the treatment qualities of RT techniques, which become more accurate but complex with time [6–7]. Via improvements in mechanical and dosimetrical QA, mechanical factors that may cause RT incidents are now controlled thoroughly [8]. However, human error is a limitation of current QA protocols, which is why it needs to be minimized. The AAPM TG-100 project is attempting to avert all possible RT incidents including those caused by human error at every step of RT procedures [9].

In conventional RT and intensity-modulated RT (IMRT), patients generally receive a high dose of radiation that is optimized in the treatment planning stage to focus on the planning target volume (PTV). To ensure that the prescribed dose is properly delivered to the PTV and a minimal dose to organs at risk (OARs), QA for RT planning is necessary. Its absence may result in under-irradiated targets, overdosed normal tissues, and higher complication probabilities. The problem is that other errors occurring in the planning phase, such as incorrect dose calculations, can be easily detected, but plan quality errors may be missed in the absence of absolute estimation standards. In current RT procedures, planners and oncologists evaluate the plan quality on the basis of experience and theory because of the difficulty of applying a uniform standard in all cases. Because these evaluations are subjective, a patient could receive a significantly different treatment result from a different medical team. Therefore, we believe that case-specific quantitative recommendations based on algorithms are needed for proper planning.

Many clinical studies have focused only on experimental factors when identifying the cause of errors in radiation treatments. Because strict mechanical QA is conducted before clinical experiments to minimize errors induced by non-experimental sources, the different results of these studies may reflect differences in RT plans [10].

Herein, we focused on plan uncertainty as a primary determinant of treatment quality. We analyzed differences in the IMRT planning quality of 42 lung cancer cases to provide preliminary data for the development of a planning QA algorithm.

## II. EXPERIMENTS AND DISCUSSION

### 1. Patient Data and Treatment Planning

To examine plan quality deviations, we collected 42 IMRT plans previously used in cases of solitary lung cancers. Multiple cancer cases were not considered. IMRT plans were created using the ECLIPSE (version 8.9) software for the radiation treatment planning system of the Varian 21iX linear accelerator. Depending on the patient's status, 5–10 fields were used and the prescribed doses were 600–6000 cGy, with averages of 4243 ± 1106, 1726 ± 363, and 1110 ± 548 cGy for the original, reduced field, and boost plans, respectively. Plans were normalized to 94.5 ± 1.0% of the prescribed dose on average. Three factors—volume ($cm^3$), length (cm), and width (cm)—were used to compare the sizes of the solitary cancers, and measurements were made using the ECLIPSE length measuring tool. Tumor volume, length, and width were 13.9–918.3 $cm^3$, 3.1–17.5 cm, and 2.8–16.5 cm, respectively. The factors determining the position of the tumor were the shortest distance from the lung apex and the shortest distance from the median line of each patient. Minimum distances from the lung apex were 0.8–10.3 cm, and minimum distances from the median line were 1.8–9.6 cm. The volume of the ipsilateral lung was 513.2–2170.1 $cm^3$, and the volume of the contralateral lung was 676.1–1982.07 $cm^3$.

### 2. Organs Considered at Risk

In lung cancer patients, organs in or near the thoracic cavity are exposed to radiation. The ipsilateral lung, contralateral lung, heart, liver, esophagus, cord, and bronchus, which are generally considered OARs during RT for lung cancer, were also considered OARs in this study. Esophagitis is a major acute complication of RT for lung cancer; it usually begins 3–4 weeks after exposure to ~30 Gy, and concurrent chemoradiotherapy may significantly increase the likelihood of esophageal injury. Other acute side effects of RT for lung cancer include coughing, skin reactions, and fatigue. Pulmonary fibrosis, esophageal stricture, cardiac sequelae, spinal cord myelopathy, and brachial plexopathy are late complications, and pneumonitis is the most common late complication [11]. The heart is the second most radiosensitive organ in the thoracic cavity after the lungs, and the rate of ischemic heart disease is directly proportional to the mean dose of radiation to which the heart is exposed [12].

In cases in which the OARs overlapped the PTV, the overlapped volume was acquired by using the ECLIPSE Boolean operator function. In non-overlapping cases, the smallest three-dimensional surface-to-surface distance between the PTV and OARs was determined using the ECLIPSE length-measuring tool.

## 3. Comparison of PTV Dose Characteristics Using Dose Volume Histograms

The coverage index (CVI), conformity index (CI), and homogeneity index (HI) of the PTV were used to compare the 42 plans [13].

$$\text{Coverage Index (CVI)} = \frac{V95_{PTV}}{V_{PTV}} \qquad (1)$$

The CVI describes the dose coverage of the PTV; $V95_{PTV}$ is the volume receiving more than 95% of the prescribed dose, and $V_{PTV}$ is a volume of the PTV [13]. CVIs ranged from 0.8 to 1 and were mostly between 0.9 and 1 (0.98 ± 0.02). There was no single CVI value, which means that targets were sufficiently covered as prescribed.

$$\text{Conformity Index (CI)} = \frac{V95_{PTV}^2}{V_{PTV} \times V95} \qquad (2)$$

The CI describes the dose conformity of the PTV; $V95$ is the volume of the whole body receiving more than 95% of the prescribed dose [14]. The closer the CI and CVI are to 1, the better the conformity and coverage. The average CI was 0.77 ± 0.08, which indicates proper conformity.

$$\text{Homogeneity Index (HI)} = \frac{|D2-D98|}{Prescribed\ Dose} \times 100 \quad (3)$$

The HI describes the uniformity of dose distribution in the PTV; *D2* and *D98* are the minimum doses received by 2% and 98% of the PTV, respectively [15]. HIs ranged from 0.11 to 0.25. The maximum value (0.25) was relatively high; lower values indicate better uniformity. HIs had a narrow Gaussian distribution (0.015 ± 0.05) in all plans except one (case no. 33), where it was 0.25. In this case, the CVI was 0.97, which indicated that the prescribed dose sufficiently covered the PTV in spite of relatively low dose uniformity. There are two reasons for the relatively low HI in this case: the shape of the PTV resembled an asymmetric dumbbell, and the PTV crossed the medial line, placing each side of the dumbbell on different sides of the lungs. Therefore, to deliver a sufficient dose to the PTV and to avoid significant damage to both lungs, most planning fields were designed to intersect the left lung, where the larger part of the PTV was located. To deliver the prescribed dose to the right side of the PTV, which consequently had fewer fields, the energy of the LAO/LPO fields was intensified to deliver prescribed dose to the right side of PTV, resulting in a high HI.

**4. Comparison of OAR Dose Characteristics Using Dose Volume Histograms**

The equivalent uniform dose (EUD), $V10_{Gy}$, and $V20_{Gy}$ of the OARs were compared in the 42 plans [13]. Measurement of EUDs facilitates comparison of treatment plans because EUDs assume a homogeneous dose distribution in the PTV and OARs when irradiation is non-homogeneous [16]. EUDs can be determined for both tumors and normal tissues. Before beginning the analysis, it was predicted that the EUDs of each OARs would be decrease as distance from the PTV increased or the OAR-PTV overlapped volume decreased.

$$\text{Effective Uniform Dose (EUD)} = (\sum_{i=1}(v_i D_i^a))^{\frac{1}{a}} \qquad (4)$$

The above formula is widely used; α is a unitless model parameter that is specific to a normal structure or tumor. The α values were 19, 13, 3, 19, and 3 for the esophagus, cord, heart, bronchus, and liver, respectively [17]. $v_i$ is a unitless value that represents the $i$'th partial volume receiving the absorbed dose $D_i$ (in Gy) when the size of each $v_i$ is 1% of the total OAR volume. $V10_{Gy}$ and $V20_{Gy}$ are the volumes receiving 10 and 20 Gy, respectively, and are used to estimate the probability of complications in normal tissues. For case no. 33, which had a relatively high HI (0.25) as described in the previous section, the EUDs for the esophagus, cord, heart, bronchus, and liver were 17.67, 6.57, 5.79, 26.06, and 0.09 Gy, respectively. These values indicate less than a 0.05 complication probability for all OARs [18].

The $V10_{Gy}$, $V20_{Gy}$, and EUDs were 28.06 ± 19.21%, 13.17 ± 14.50%, and 8.06 ± 5.10 Gy for the ipsilateral lung and 6.53 ± 8.89%, 1.18 ± 2.41%, and 2.59 ± 2.01 Gy for the contralateral lung, respectively. As the lungs are the most radiosensitive organ in the thoracic cavity, they must be managed carefully to avoid pneumonitis [19]. The complication probabilities of the lungs were 0.05 (~100% of the volume) for the $V10_{Gy}$ and 0.05 (0–40% of the volume), 0.1–0.2 (~60% of the volume), and >0.5 (~100% of the volume) for the $V20_{Gy}$ [17]. Most cases had less than a 0.05 complication probability, although one case (no. 26) had a $V20_{Gy}$ >0.4 and a complication probability between 0.1 and 0.2. $V10_{Gy}$ and $V20_{Gy}$ were relatively high in cases with large PTVs, and EUDs tended to increase as the PTV increased (Fig. 1). The ipsilateral lung showed a discernible increasing trend, whereas the tendency of the contralateral lung was unclear. The heart had a complication probability of less than 0.05 on average, and the cord, liver, and esophagus had a complication probability of less than 0.01 on average (Table 1) [19].

Among the 42 cases, the esophagus, cord, heart, bronchus, and liver were separated from the PTV in 22, 41, 19, 11, and 38 plans, respectively. EUDs showed a slight downward tendency as the CI decreased and the distance from the PTV increased in most OARs, as predicted (Fig. 2). PTVs overlapped the esophagus, cord, heart, bronchus, and liver in 20, 1, 23, 31, and 4 plans, respectively. The EUDs of the bronchus and heart tended to increase as overlap volume increased (Fig. 3). As the cord and liver are not in the thoracic cavity, they were separated from the PTV in most cases, making it difficult to discern general tendencies. The EUDs of the esophagus barely showed a tendency; as the esophagus has a relatively high tolerance dose (<0.01 complication probability for the whole esophagus in the $V30_{Gy}$), planners usually optimize the target dose to avoid pneumonitis rather than minimize the esophagus dose in cases where the PTV overlaps the esophagus.

### III. CONCLUSION

In the absence of a standard QA protocol for RT planning, substantial discrepancies in plan quality in similar cases are inevitable. Establishing quantitative references, such as algorithms, for treatment planning can considerably reduce deviations in treatment quality between planners. Most of the 42 plans we examined followed the expected tendency (i.e., EUDs of OARs decrease as distance from the PTV increases or OAR-PTV overlapped volume decreases) except for one, which had a relatively high HI (0.25, case no. 33). However, singular values do not always indicate subquality of the plan because minimizing the doses to specific OARs may not receive priority if they do not exceed the critical tolerance dose and there are other considerations such as pneumonitis. Consequently, the focus and subjective decisions of the planners play a significant role in complex cases. Case-specific planning algorithms may represent the best option in difficult cases, although there are no absolute solutions to the problems of RT planning.

Because the plans in this study were from a single department, relatively few people were involved in treatment planning. The differences in the treatment results for a given patient would be much more pronounced if multiple departments (and therefore more planners) were involved. Therefore, further examination of QA protocols is needed to reduce deviations in radiation treatment planning.

## ACKNOWLEDGEMENT

This work was supported by the General Researcher Program (NRF-2012R1A1A2003174), the Nuclear Safety Research Program (Grant No. 1305033 and 1403019) of the Korea Radiation Safety Foundation and the Nuclear Safety and Security Commission, Ministry of Food and Drug Safety (Grant No. 1517NIFDS395 ) and Radiation Technology Development Program (2013M2A2A4027117) of the Republic of Korea.

Table 1. The equivalent uniform dose (EUD), $V10_{Gy}$, and $V20_{Gy}$ of the organs at risk

| Site | EUD (Gy) | $V10_{Gy}$ | $V20_{Gy}$ |
| --- | --- | --- | --- |
| Esophagus | 6.86 ± 5.44 | 25.53 ± 20.37 | 12.02 ± 16.64 |
| Cord | 3.48 ± 3.17 | 13.23 ± 17.23 | 4.37 ± 9.17 |
| Liver | 7.03 ± 1.73 | 2.10 ± 6.32 | 0.89 ± 2.94 |
| Heart | 5.03 ± 5.32 | 19.20 ± 25.34 | 8.41 ± 14.35 |
| Bronchus | 10.33 ± 11.38 | 35.20 ± 29.59 | 19.31 ± 27.66 |

Fig. 1. Equivalent uniform doses (EUDs) of the ipsilateral lung and contralateral lung as a function of the planning target volume (PTV).

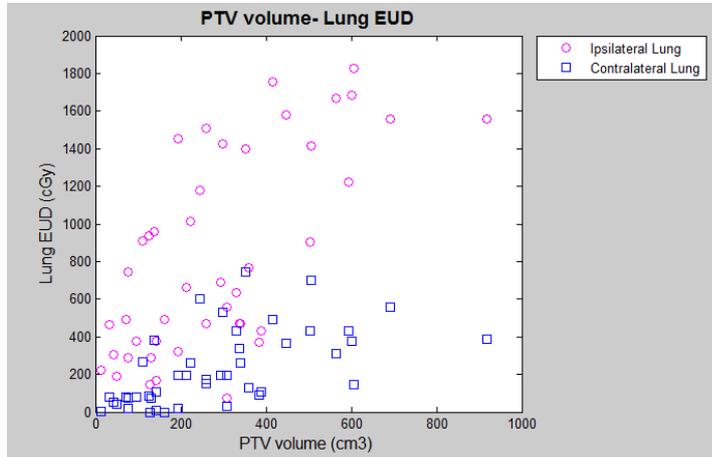

Fig. 2. Equivalent uniform doses (EUDs) of the organs at risk according to the conformity index and distance (cm) from the planning target volume. (a) Esophagus; (b) Cord; (c) Heart; (d) Bronchus; (e) Liver

(a) 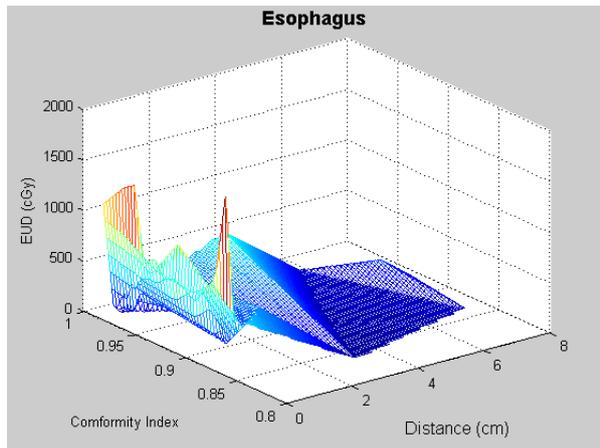 (b) 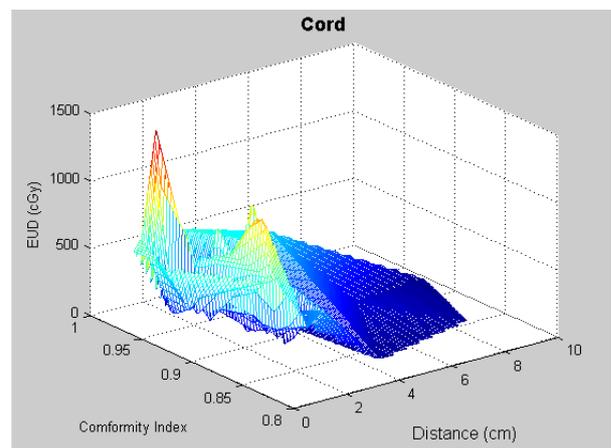

(c) 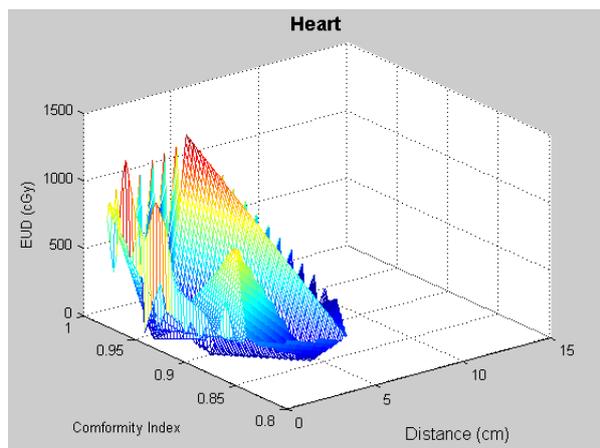 (d) 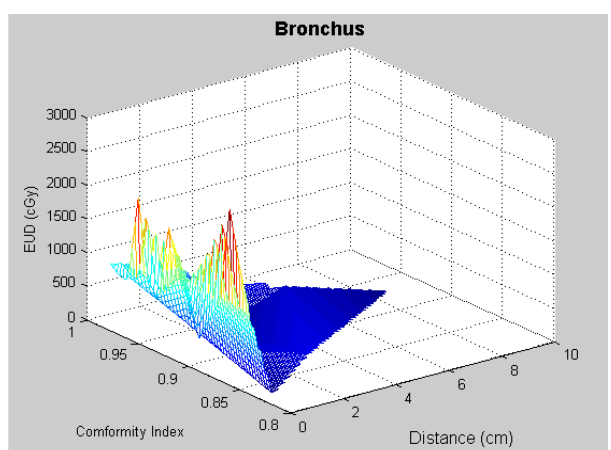

(e) 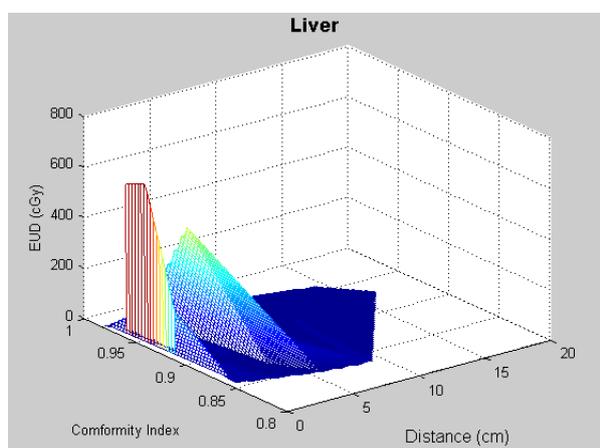

Fig. 3. Equivalent uniform doses (EUDs) of the organs at risk according to their overlap (cm$^3$) with the planning target volume. (a) Esophagus; (b) Cord; (c) Heart; (d) Bronchus; (e) Liver

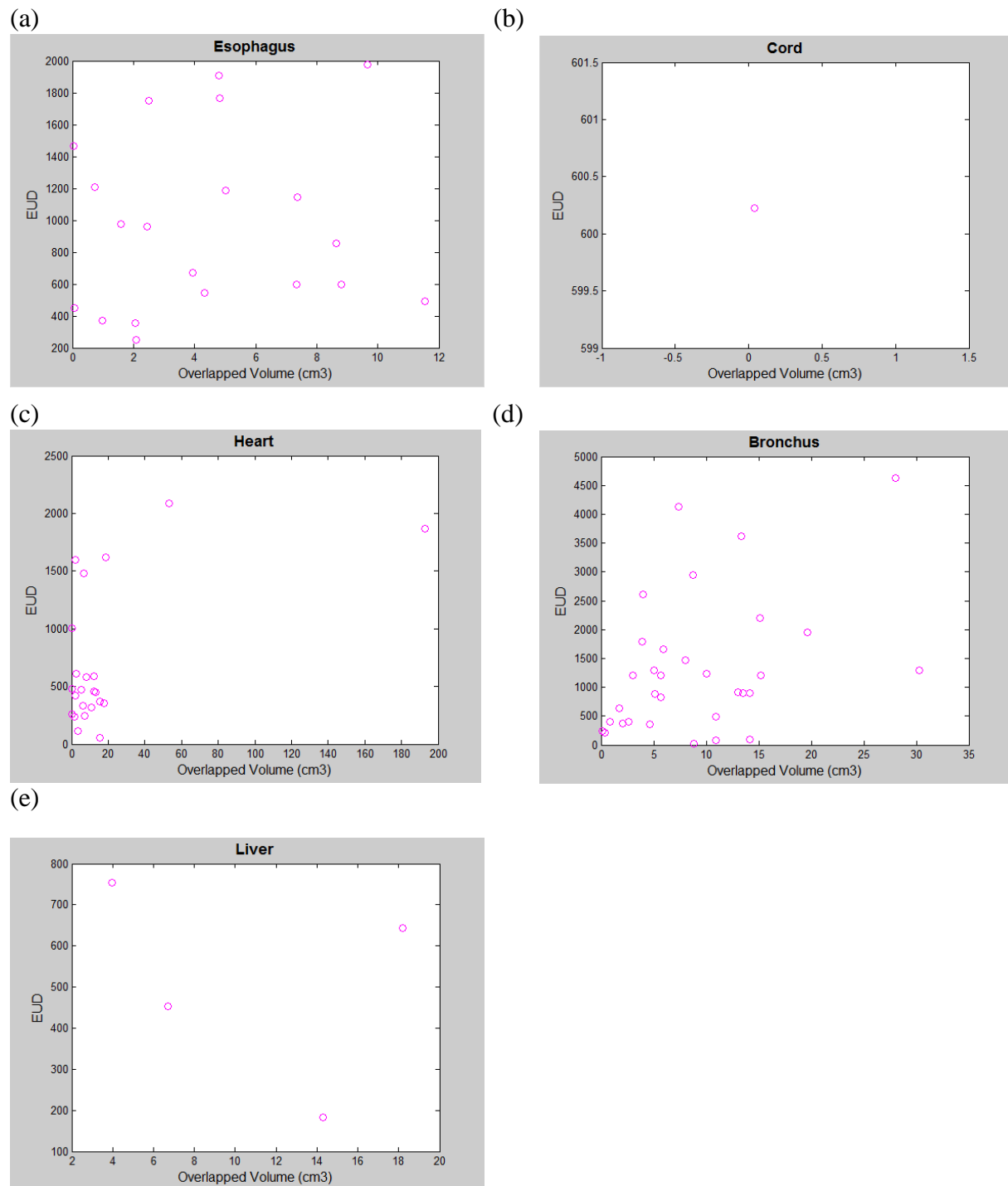